\documentclass[12pt,a4paper]{article}
\usepackage{a4p}
\usepackage{cite,mcite}
\usepackage[dvips]{graphicx}
\usepackage[dvips]{epsfig}
\usepackage{physics }
\usepackage{l3_title,ifthen}
\journalname{NIM}
\date{March 27, 1997}
\preprint{PREPRINT DFF 278/4/1997}
\graphicspath{{/l3/paper/example/figs/}}
\newlength{\capindent}
\newlength{\capwidth}
\newlength{\figwidth}
\newcommand{\icaption}[2][!*!,!]{\hspace*{\capindent}%
  \begin{minipage}{\capwidth}
    \ifthenelse{\equal{#1}{!*!,!}}%
      {\caption{#2}}%
      {\caption[#1]{#2}}
  \end{minipage}}
\begin{document}
\begin{titlepage}
\title{Confidence level estimation and analysis optimization}

\author{A. Favara and M. Pieri}

\vspace{1.5cm}

\author{INFN and University Firenze}
%
% The abstract
%
\begin{abstract}

This note proposes a method,
which can be applied to searches and more in general
to any cross section measurement, to maximize the
analysis sensitivity.

\end{abstract}
\end{titlepage}

\def\EVIS{\ensuremath{E_{\mathrm{vis}}}}%
\def\ETRA{\ensuremath{E_{\perp}^{rel}}}%
\def\EPAR{\ensuremath{E_{\parallel}}}%
\def\EV30{\ensuremath{E_{\mathrm{v30}}}}%
\def\NATK{\ensuremath{N_{\mathrm{tk}}}}%
\def\GDTK{\ensuremath{N_{\mathrm{gtk}}}}%
\def\NASR{\ensuremath{N_{\mathrm{cl}}}}%
\def\rphi{\ensuremath{R-\phi}}%
\def\Ebar{\ensuremath{E\hspace{-.23cm}/\hspace{+.01cm}}}
\def\EM25{\ensuremath{\Ebar_{25}}}
\def\EMF25{\ensuremath{\Ebar^{\perp}_{25}}}
\def\ECM60{\ensuremath{\Ebar^b_{60}}}
\def\PORT{\ensuremath{E_{\perp}}}
\def\THPM{\ensuremath{\sin(\theta_{miss})}}
\def\NMPP{\ensuremath{\Delta\phi_{tk}}}
\def\XNUOVA{\ensuremath{E_{TTL}}}
\def\PTRAL{\ensuremath{P_{\perp l}}}
\def\XNUOVAJ{\ensuremath{E_{TTJ}}}
\def\TKM25{\ensuremath{N_{tk}^{25}}}
\def\TRE{\ensuremath{\theta_{123}^l}}
\def\TREP{\ensuremath{\phi_{123}^l}}
\def\ELEP{\ensuremath{E_{\mathrm{lep}}}}%
\def\ELUM{\ensuremath{E_{\mathrm{LUM}}}}%
\def\EALR{\ensuremath{E_{\mathrm{ALR}}}}%
\def\EBGO{\ensuremath{E_{\mathrm{BGO}}}}%
\def\EHCL{\ensuremath{E_{\mathrm{HCL}}}}%
\def\ENMU{\ensuremath{E_{\mathrm{MUCH}}}}%
\def\ETAU{\ensuremath{E_{\mathrm{\tau}}}}%
\def\EGAM{\ensuremath{E_{\mathrm{e\gamma}}}}%
\def\OPTI{\ensuremath{\ast}}
\def\PORTL{\ensuremath{E_{\perp}^{lep}}}
\def\MAXACO{\ensuremath{\Delta\phi_{lep}}}
\def\TTHPM{\ensuremath{\theta_{miss}}}
\def\DELTAM{\ensuremath{\Delta{\mathrm{M}}}}%
\def\CL{\ensuremath{CL}}%
\def\PL{\ensuremath{{\cal P_L}}}%
\def\PT{\ensuremath{{\cal P}_T}}%
\def\PLV{\ensuremath{\bar{\cal P_L}}}%
\def\PTV{\ensuremath{\bar{\cal P}_T}}%

\def\simge{\ \lower -2.5pt\hbox{$>$} \hskip-8pt \lower 2.5pt \hbox{$\sim$}\ }
% definizione di minore circa
\def\simle{\ \lower -2.5pt\hbox{$<$} \hskip-8pt \lower 2.5pt \hbox{$\sim$}\ }

%%%%%%%%%%%%%%%%%%%%%%%%%%%%%%%%%%%%%%%%%%%%%%%%%%%%%%%%%%%%%%%%%%%%%%%%%%%%%%%
% Introduction
%%%%%%%%%%%%%%%%%%%%%%%%%%%%%%%%%%%%%%%%%%%%%%%%%%%%%%%%%%%%%%%%%%%%%%%%%%%%%%%
%
\section{Introduction}
During the last years at LEP a new attitude toward searches 
has developed which tends to separate the actual search 
for new phenomena from the derivation of an exclusion limit.
In this note we would propose a new method of globally optimized
analysis which
combines the two steps in an optimal and unbiased way by
making use of background subtraction.

\section{Statistical Issues}\label{statiss}
Several methods have been proposed at LEP
for the extraction of limits on the Higgs boson mass \cite{obraz, grivaz, bock}
and relative selection optimization \cite{[3]},
while for most of other searches the limits are based 
on the PDG formula \cite{pdghelene}.
No methods, to our knowledge, have been proposed yet
to optimize an analysis aimed at discovery
and deriving the related discovery confidence level (\CL).

When measuring physical quantities 
the combination of different results with gaussian errors is straightforward.
On the other hand it is usually more complicated to
combine different exclusions or discovery \CL.

Clearly to derive a discovery \CL\ the background has to be subtracted.
Therefore, aiming at a globally optimized analysis
capable of discoveries and exclusions at the same time
we should definitely consider background subtraction.

To make a \CL\ evaluation there are two possibilities:
\begin{itemize}
\item either one has a formula capable of calculating the \CL\ of
the outcome of a given experiment
\item or one defines an estimator to rank the experiments
from the more signal(background)--like to the less signal(background)--like
and then, either analytically or by use of Monte Carlo (MC) techniques,
evaluates the ranking of the actual experiment.
\end{itemize}

Some formulae have been proposed for the first case
as some \CL\ evaluation methods have been proposed for the second case.
We would like to note that in the second case the estimator
can be built arbitrarily (provided that it is done a-priori).
%Also in some cases the analysis is performed without background
%subtraction and therefore is not suited to be used for discovery.

In all these approaches the only thing which is used to derive
the \CL\ is the signal to background ratio in the points where the candidates
are observed.

The formula in \cite{obraz}, derived in the Bayesian approach,
is very simple and it naturally includes background subtraction
and therefore in the following we will use it.
As it has not yet been demonstrated that
this formula gives a statistically well defined \CL\
we use its output as an estimator which will then be
treated with MC methods.
It should be noted at this point that the only difference
between the methods is due to the different estimator chosen.
Our choice, which is anyway not essential for the results of the method,
 is due to the simplicity and naturalness of the 
estimator described in \cite{obraz} and also to the fact that its output
may sometimes be a little conservative but it is a very
good approximation of the \CL.

The MC method has on the other hand the big advantage that 
it allows, with virtually no additional effort, to include
any sort of statistical and systematic errors on efficiency
and background when computing the \CL.

\section{Confidence Levels}

For sake of clarity we try in this section to explain in simple words
the definition
of confidence level and the different methods used to derive it.

The results of a measurement are compared with a given hypothesis,
for instance that the Higgs boson has a given production
cross section, and
the confidence level associated to the exclusion of our hypothesis
gives the probability that the disagreement
between the measurement and the  given hypothesis is smaller than
what actually observed.

As previously stated we can either do an analytical calculation to derive
the confidence level for a given experiment, or we can evaluate it with
Monte Carlo technique. Unfortunately the first possibility seems not to be
viable because it is virtually impossible to obtain
an analytical expression to relate any given
estimator, that we want to use to rank the experiment in the most 
effective way, with its probability density function. 

The cross check of the analytical calculation of the confidence
level can be done by means of Monte Carlo simulations of trial
experiments. In practice this means that we perform 
our measurements many times, introducing each time a number of background
and signal events according to a Poisson distribution with mean
equal to the sum of the expected background and signal (our hypothesis) levels.

The actual probability associated to the calculated confidence level
is then simply given by the rate of simulated measurements in larger 
disagreement with the  given hypothesis than the real one,
properly taking into account the fact that the background alone
must give a value of the estimator smaller than the observed one.
The latter means that the background can not fluctuate over the actual
number of events observed in the experiment and that its 
probability distribution is bound from above. 
If we do not take into account that, as already reported in Reference
 \cite{pdghelene}, we can derive the absurd conclusion that every signal,
even with null cross section, can be excluded with extreme over-fluctuation
of the background.

It is clear that we can use this  Monte Carlo cross check directly
as a definition of the confidence level. This turns out to be a powerful
tool because we can use what we believe the most sensitive estimator to rank
the measurement and then the probability associated to it is simply
obtained as previously explained.

\section{The Estimator}

We would first of all like to stress that the estimator we used \cite{obraz} 
is a very simple extension of the PDG prescription
to compute \CL\ for experiments with or without background subtraction 
\cite{pdghelene} to the multichannel situation:
$$F(s,b,n) = 
e^{-(s+b)}\prod^{k}_{j=1}\frac{(s\cdot f_j + b\cdot g_j)^{n_j}}{n_j!}$$
where the  index $j$ runs over the different channels and where
$f_j$ and $g_j$ are the fraction of all signal events ($s$) and all
background events ($b$), which are in the channel $j$. 
The fraction of the total number of observed
candidates $n$, in each channel $j$, is $n_j$.
Using the Bayesian approach we can obtain the confidence level
for a given number of expected signal events $\sigma$ 
with the following integrations:
\begin{equation}
\PL = \frac{\int_{\sigma}^{\infty} F(s,b,n) ds}
{\int_{0}^{\infty} F(s,b,n) ds} \simeq 1 - \CL\
\label{confi}
\end{equation}
\PL\ is the estimator we use to rank the experiment.

Using it as an estimator, it is straightforward to
extend it to be without background subtraction. It is enough
to compare the estimator with the distribution
of signal-only random experiments.

To get the correct \CL\ we compare the actual 
value of the estimator
to the distribution of trial MC experiments using signal and background.
Clearly, given the fact that the experiment excludes
the possibility of having a background which gives by itself
a larger estimator than the measured one we only consider
those MC experiments with the background estimator smaller
than the measured estimator. We add to those a random signal,
then measure the value of the probability.
Usually, as shown below, the true \CL\ is slightly different
from the one we can obtain from equation \ref{confi}.
The \CL\ calculated with equation \ref{confi},
for a small number of events, is always conservative.
For a large number of events,
it tends to be a correct estimation of the actual \CL\
measured with MC experiments.

\section{Analysis Optimization}

As said in Section \ref{statiss} all methods for limits calculation
proposed for a single channel are only sensitive to the 
signal/background ratio
in the point or analysis (or whatever) where the data lie.

The new method proposed here is to consider this fact
and its direct consequence that an intrinsically
optimized analysis can be obtained 
when for any value of the ratio signal/background we obtain the highest 
possible overall efficiency.

It is important to understand that the specific power of 
a subanalysis only depends on the signal to background ratio and not on
the absolute amount of the signal selected by the subanalysis.
Then the optimization can be achieved including all the
available information provided that channels/analyses with different
signal/background ratio are not mixed. This means that we have to
organize a set of complementary analyses (i.e. each one selects events
that no other analysis selects) and then whatever analysis we
add the overall sensitivity is improved.

If we are able to devise such an optimal selection,
which orders the different subselections in terms of
signal to background ratio, whatever
estimator we use, the analysis can be optimized by
a single cut on the signal to background ratio.
The latter in practice is often needed to take into account
systematic uncertainties.

\begin{figure}[hbtp]
\begin{center}
\begin{tabular}{cc}
\mbox{\epsfysize=9.0cm \epsffile{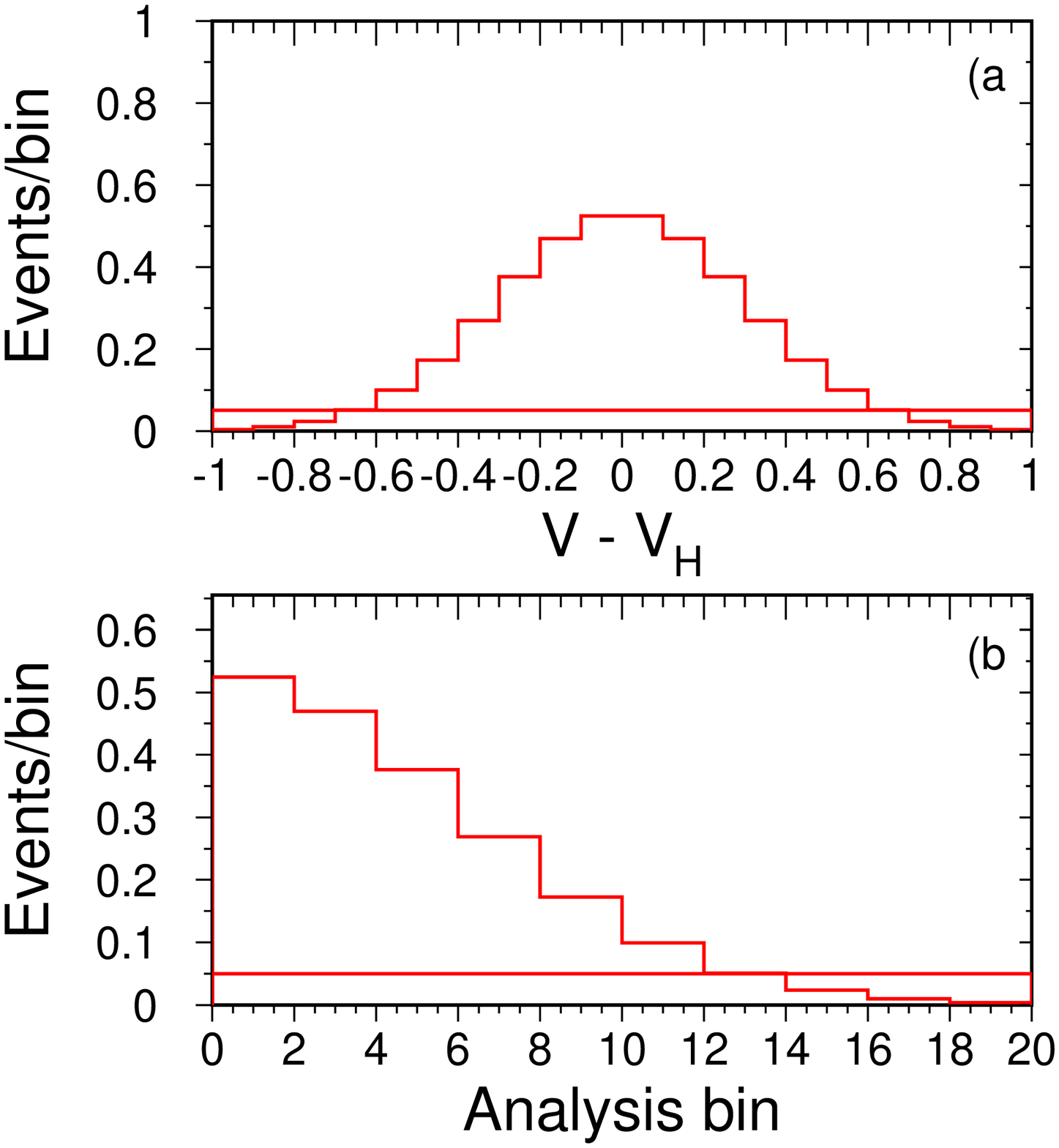}} &  
\mbox{\epsfysize=9.0cm \epsffile{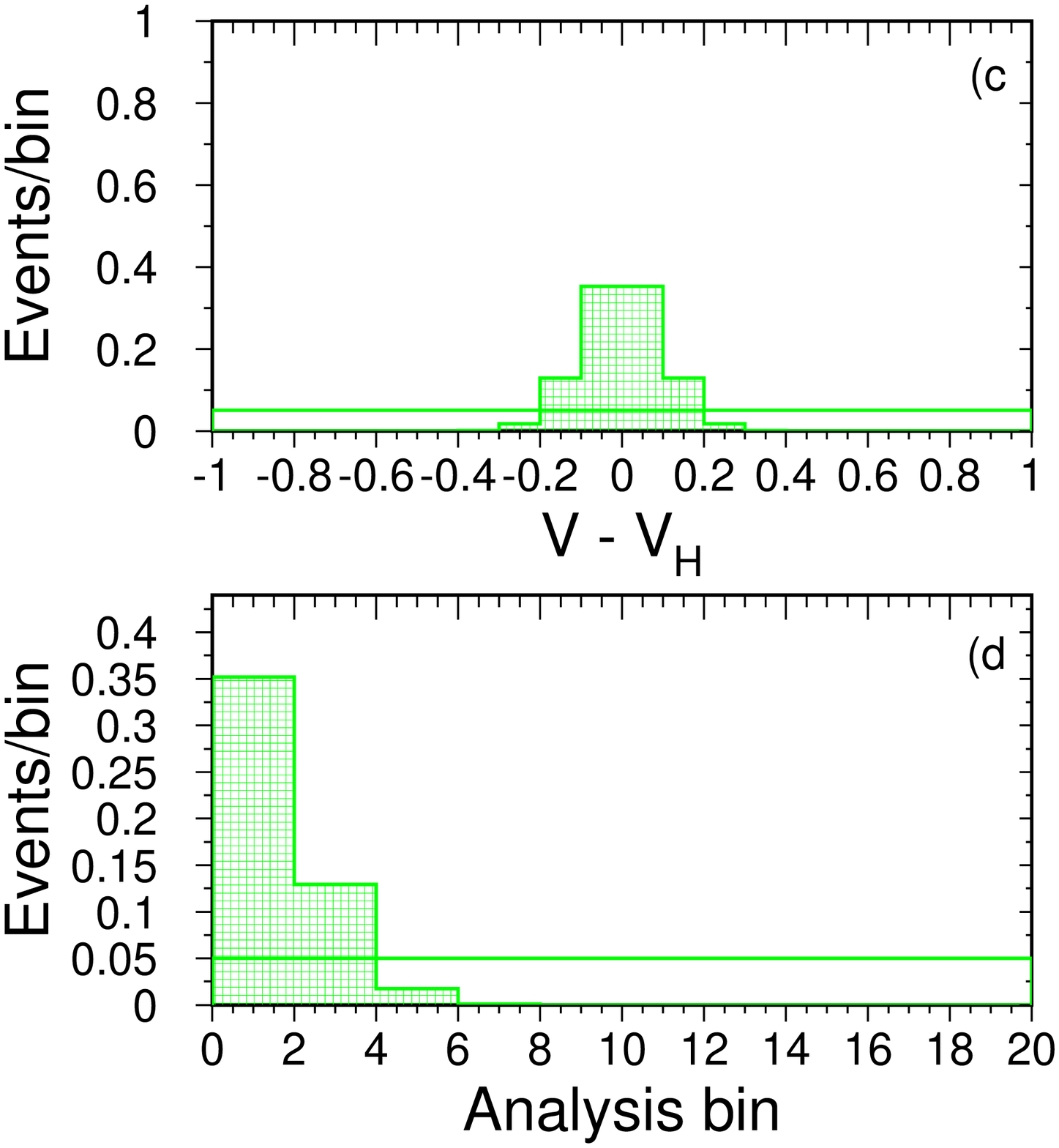}} \\
\mbox{\epsfysize=9.0cm \epsffile{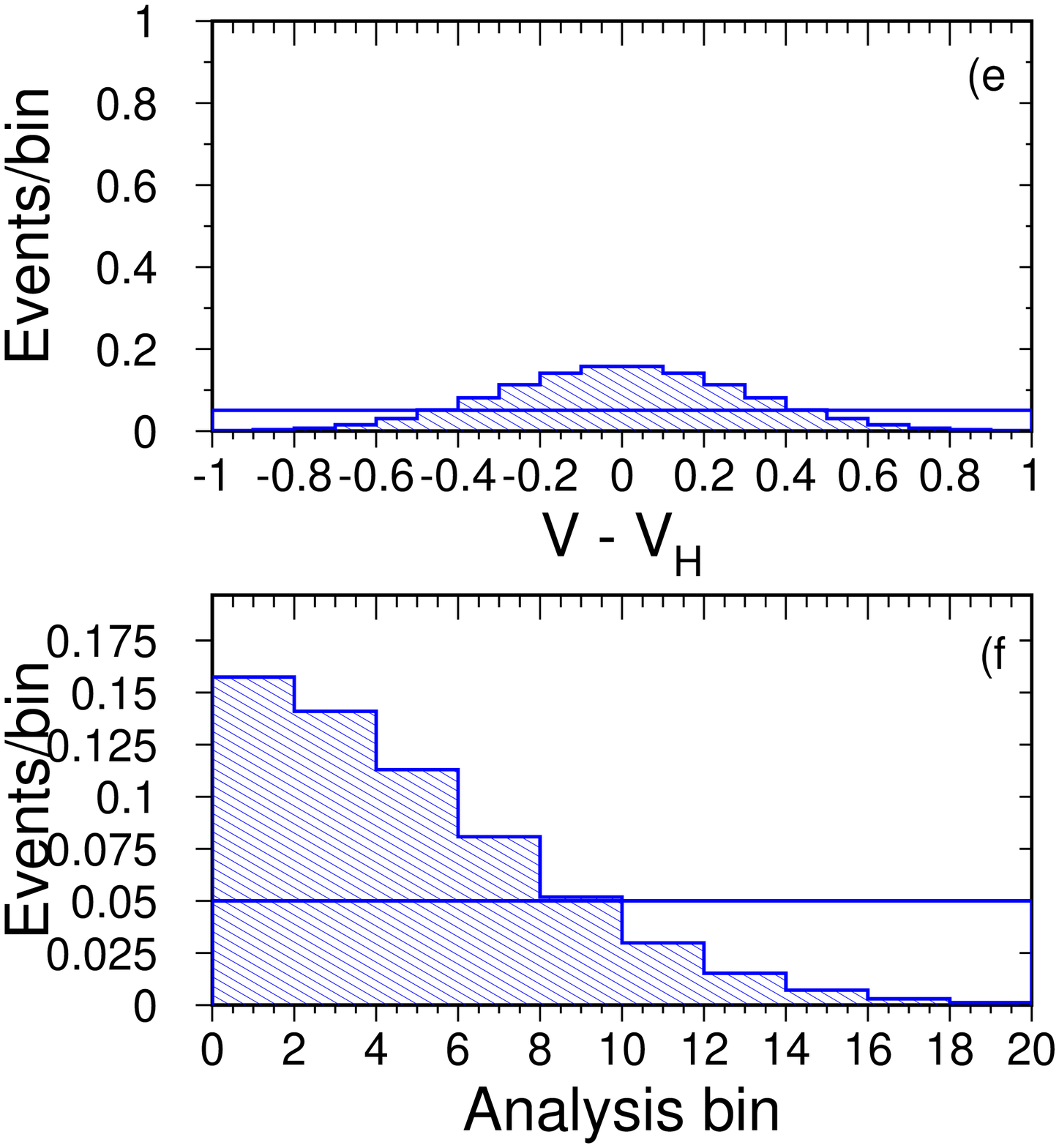}} &
\mbox{\epsfysize=9.0cm \epsffile{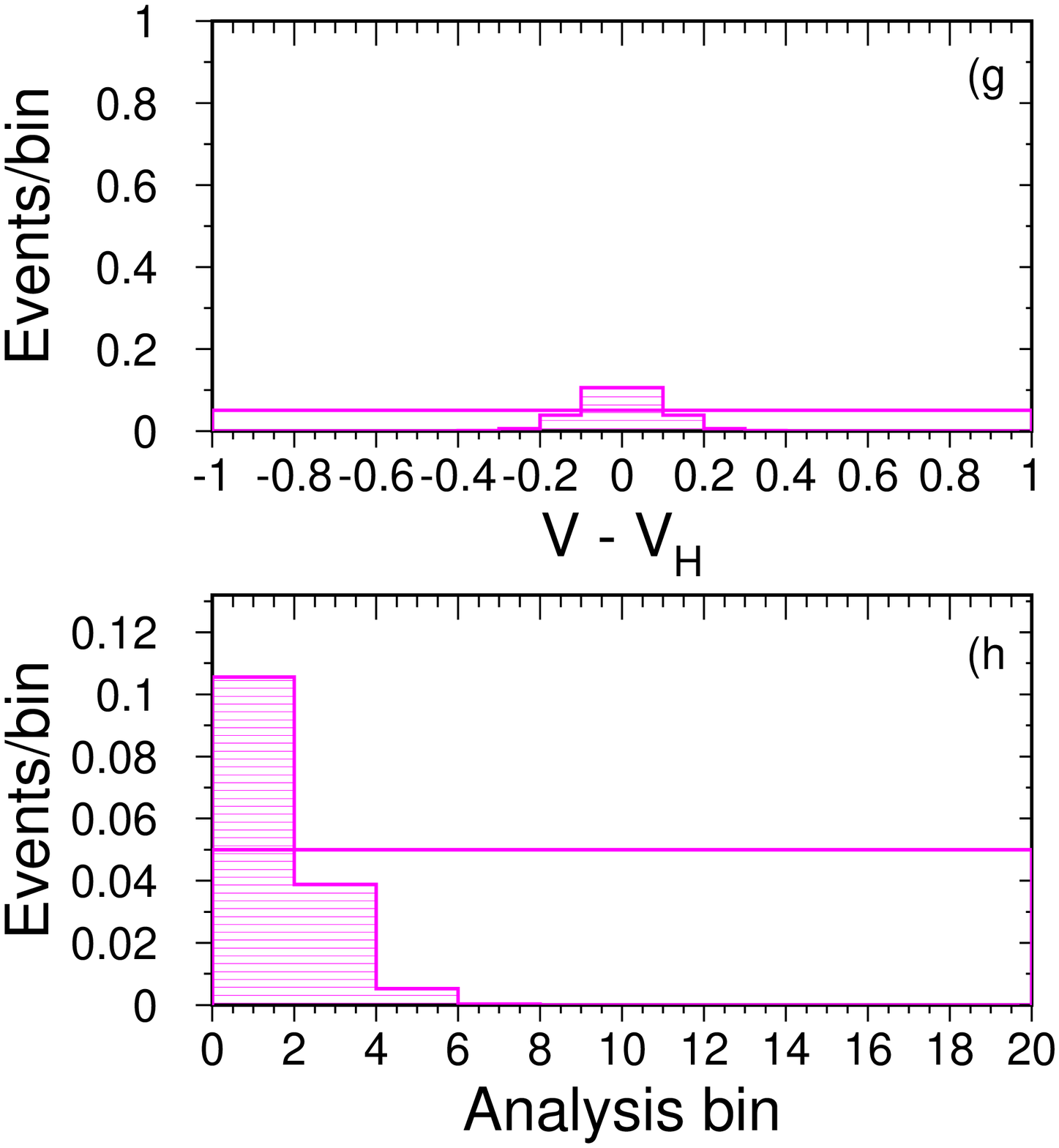}}
\end{tabular}
\caption{\label{exanf} 
Distributions for signal and background in analyses
A (a) B (c) C (e) D (g). Equivalent distributions,
analyses A (b) B (d) C (f) D (h),
with signal/background ratio ordering.
}
\end{center}
\end{figure}

To illustrate these concepts, let us consider a simple numerical example.
Let us take the usual case of the search for the production of a
new particle which can occur in two channels A and B
with branching ratios of 20\% and 80\% respectively.
Let us assume that in the two channels
we have the same selection efficiency of 50\% with a mass resolution
for the signal of 0.3 in channel A and 0.1 in channel B,
and that we have a flat background of 1 event from --1 to 1
(we take as an example the mass but it can be any other variable).
Binning such a distribution in bins of width of 0.1,
the result is shown in figure \ref{exanf}-a for channel A and
\ref{exanf}-c for channel B.

\begin{figure}[hbtp]
\begin{center}
\mbox{\epsfysize=12.0cm \epsffile{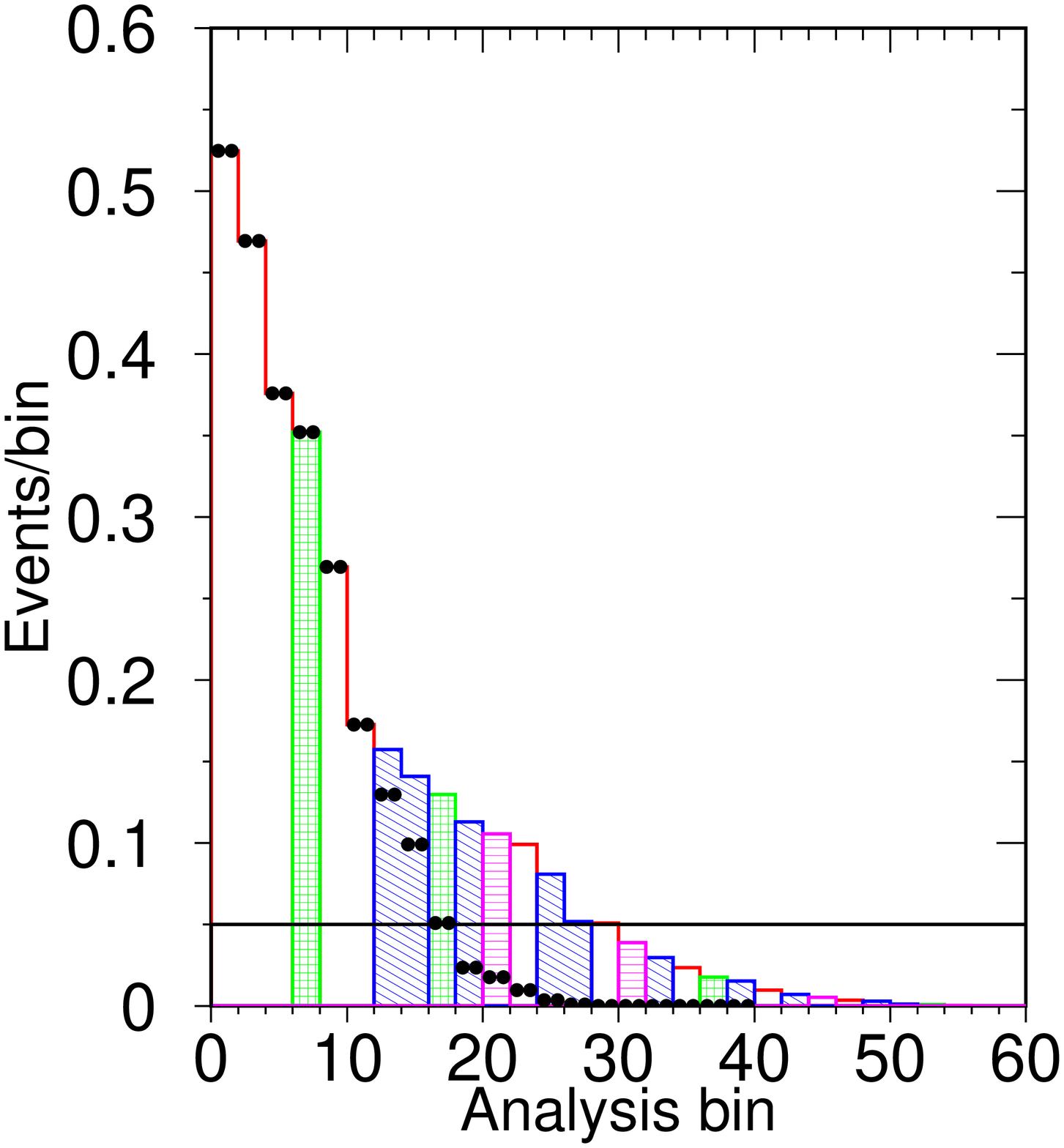}} 
\caption{\label{exant} 
Combination of analyses A B C D (histogram) and of analyses
A and B only (dots). Analysis bins are ordered according to their
signal/background ratio.}
\end{center}
\end{figure}

 These two spectra can be rearranged as in figure \ref{exanf}-b 
for channel A and \ref{exanf}-d for channel B,
 where the bins of the two analyses are ordered in
 signal/background ratio.
 The combination of the first two analyses, shown in figures \ref{exanf}-b
and \ref{exanf}-d, is represented by the dots in figure \ref{exant}. 

At this point if we increase the signal efficiency we can improve
the overall sensitivity.
 For instance we take two additional selections, C and D,
 both with 15\% additional efficiency
 on channel A and B respectively. For simplicity the mass resolution 
and the background level for the new analyses 
are assumed to be the same as those in the original analyses A and B.
 The equivalent plots are also shown in figures \ref{exanf}-e-f
for analysis C and in figures \ref{exanf}-g-h for analysis D.

From the combination of the four analyses, shown in figure \ref{exant},
 it is apparent that the A-B-C-D analyses combined are better
 than A-B analyses alone.
 As it will be explained in more detail in the following it is clear that,
 depending on the estimator used and on the systematic errors, it may be
 useful to reject the bins with worse signal/background ratio.
 In conclusion, to obtain a globally optimized analysis for exclusion and
 discovery  at the same time it is sufficient
 to have an optimized ordering of the analyses, or of the analysis bins,
 which can be  obtained with the best available variables.

LEP experiments so far for the calculation of Higgs limits have
used estimators which evaluate the signal/background ratio
by means of the reconstructed Higgs mass. This is not always
the best information to discriminate signal from background (for instance
in the \Ho\qqbar\ channel the b--tagging is much better) and
in any case is not the only one.\\
One possibility, for example, is to use neural network and the neural network
output. Another is to use simple selections cuts linked together by means
of a global variable as previously shown in \cite{l3_102}.
The details of this last technique
will be described in a dedicated paper \cite{supap}.

According to formula  \cite{pdghelene}, where
background subtraction is used, adding a new channel, 
even with very poor signal/background ratio, 
the average analysis sensitivity does not worsen.

This statement implies that all channels should be used
because they improve the overall sensitivity, even if very little.
Unluckily this is an ideal situation because, as we will see
in the following, uncertainties on the background
level, both from statistical and systematic sources, do not
allow the use of analyses with very poor signal/background ratio.
So the optimal analysis should have a cut on the signal/background ratio
according to the amplitude of the uncertainties.\\
In particular as we have already seen it is better
not to join in a single analysis channels with different 
signal/background ratio.
Furthermore we split every single channel/analysis into
several complementary analyses with different signal/background ratios, and
we try to do that in the most effective way. But
we would like to warn anyway that one should be cautious with
that procedure because it may lead, if we do not take into account
uncertainties on the background level, to average and sometimes also actual
limit overestimation due to large MC statistical errors in the different 
bins of the analysis. On the other hand if we take into account properly the
uncertainties on the background level and these are too high, we are then
obliged to apply a tight cut on the signal/background ratio.\\
Concluding the definition of the optimal analysis also depends
on the amount of MC systematic uncertainties.

To better explain this second important issue of our method, i.e.
how many analyses and how much background should be accepted,
we will treat in the following
section a simple example, rather than describing the application of
this method to Higgs searches at LEP $161\div 172$ as reported in
\cite{notah172}.

\section{A Simple Example}

\begin{figure}[hbtp]
\begin{center}
\mbox{\epsfysize=9.0cm \epsffile{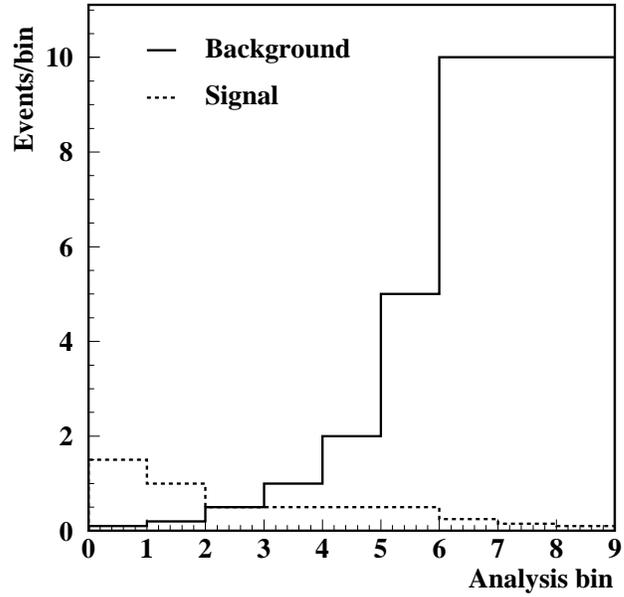}} 
\caption{\label{figexa} Number of expected events for signal 
and background for the different analyses. For the signal we assumed
5 events expected.}
\end{center}
\end{figure}

Let us now make a simple example, let us consider the case 
of a binned analysis with different efficiencies
and background contaminations.
The assumed efficiencies and background are reported in table \ref{tabexa}
and the number of expected events for signal and background are 
shown in figure \ref{figexa}.
\begin{table}
  \begin{center}
    \begin{tabular}{|c|c|c|c|}\hline
Analysis & Efficiency & Expected background & S/B\\
\hline
1 &    0.3     &    0.1& 15.000\\
2 &    0.2     &    0.2& ~5.000\\
3 &    0.1     &    0.5& ~1.000\\
4 &    0.1     &    1.0& ~0.500\\
5 &    0.1     &    2.0& ~0.250\\
6 &    0.1     &    5.0& ~0.100\\
7 &    0.05    &   10.0& ~0.025\\
8 &    0.03    &   10.0& ~0.015\\
9 &    0.02    &   10.0& ~0.010\\
\hline
    \end{tabular}
    \caption{Efficiency, number of events expected from the 
background and signal/background ratio for nine different analyses.
   \label{tabexa}}
  \end{center}
\end{table}
Each bin can be considered for example as a different decay channel
of the same physical process (for instance Higgs production).

\subsection{Statistical errors only}

First of all we consider the case of background subtraction with no 
systematic errors to approach the problem only on the statistical side.

\begin{figure}[hbtp]
\begin{center}
\begin{tabular}{cc}
\hspace{-1cm}
\mbox{\epsfysize=9.0cm \epsffile{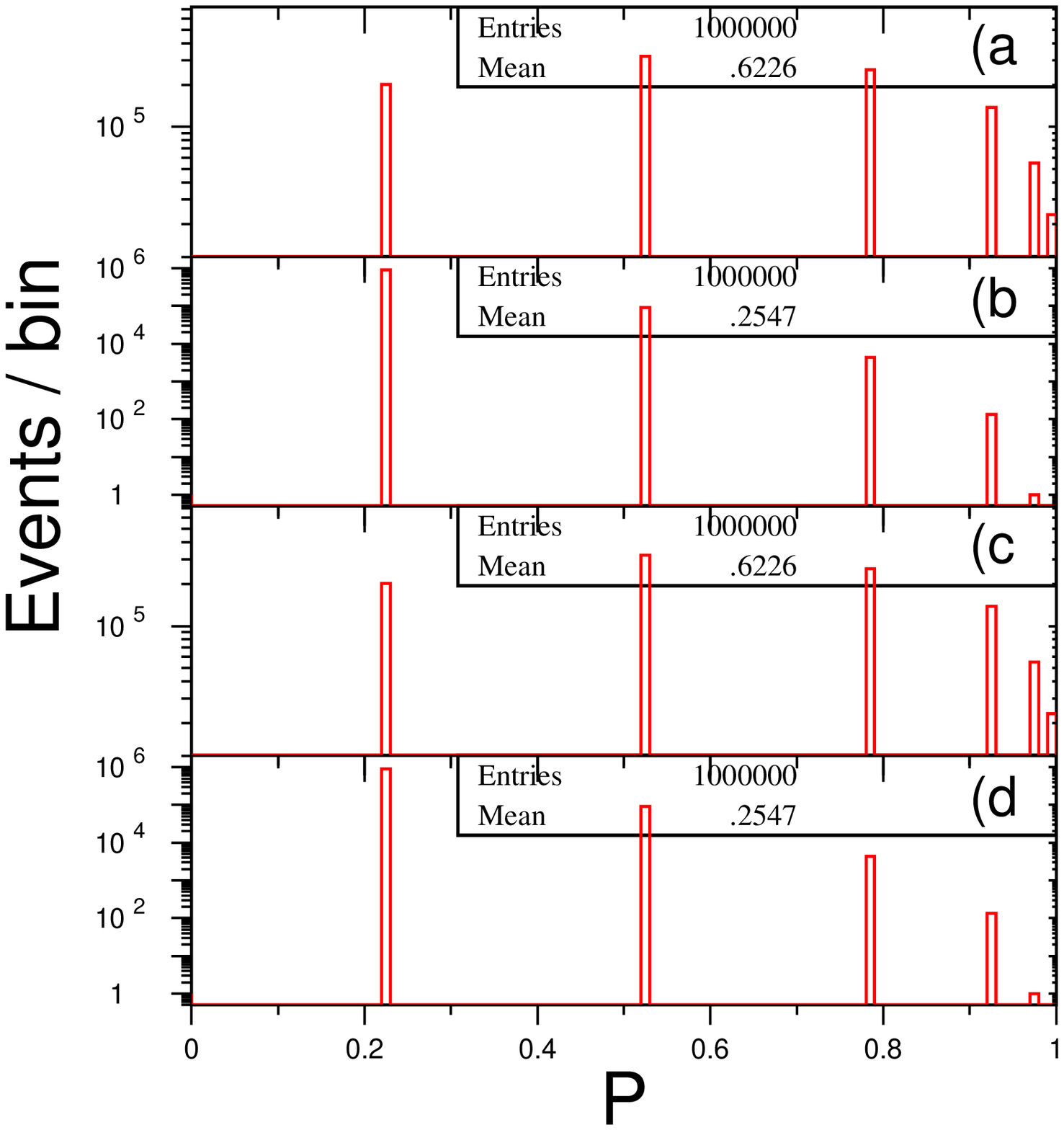}} &  
\mbox{\epsfysize=9.0cm \epsffile{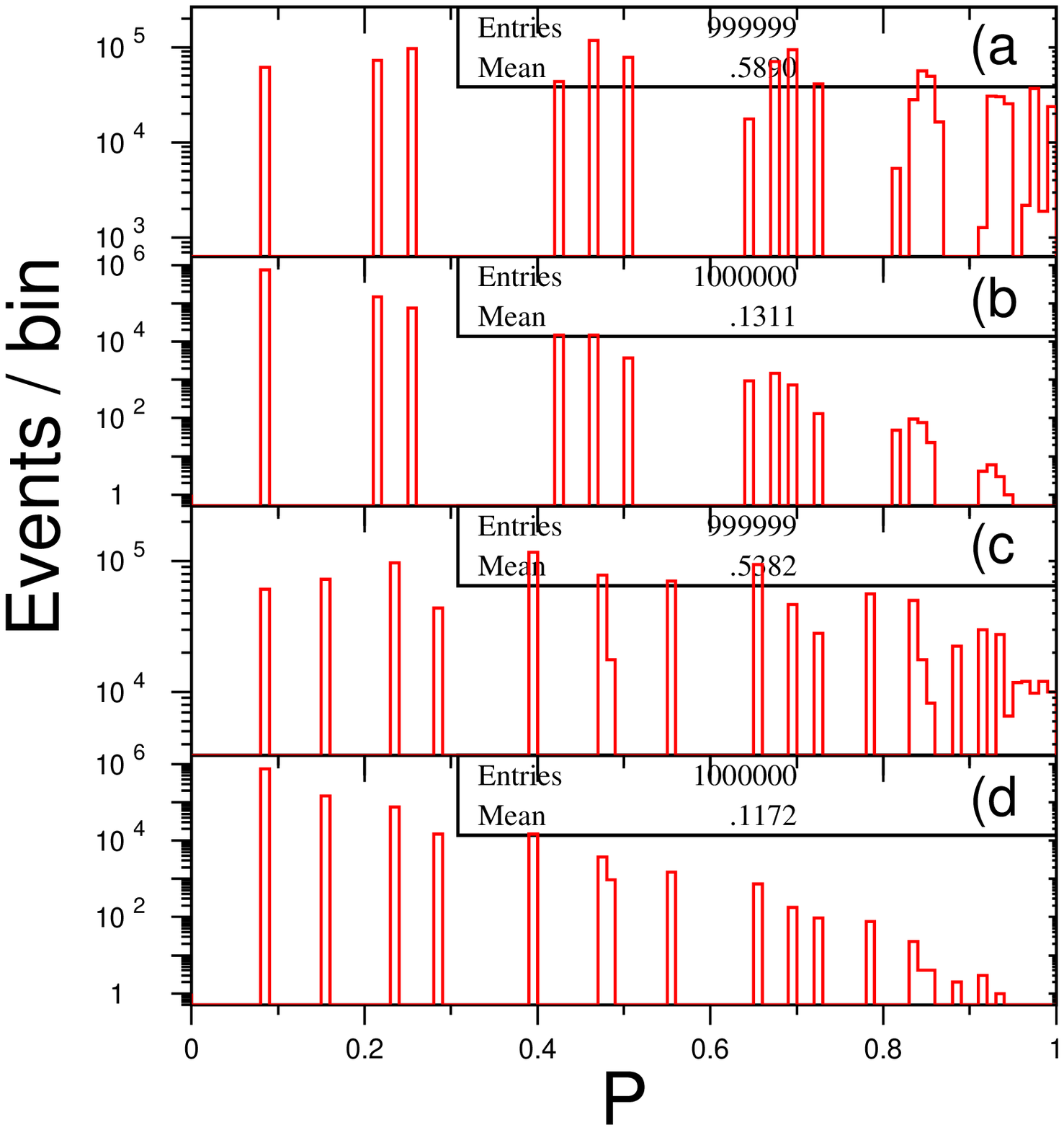}} \\
\hspace{-1cm}
\mbox{\epsfysize=9.0cm \epsffile{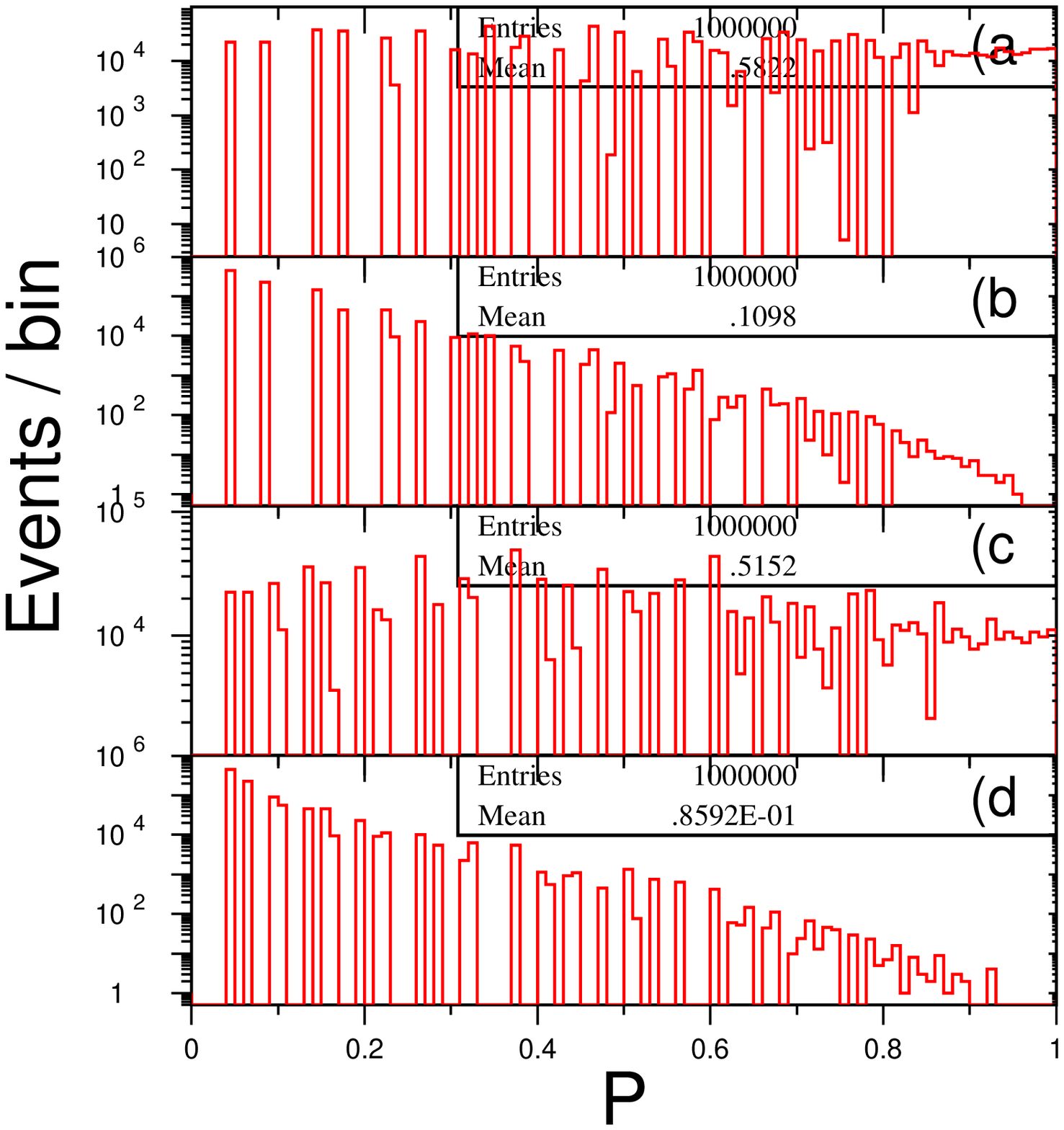}} &
\mbox{\epsfysize=9.0cm \epsffile{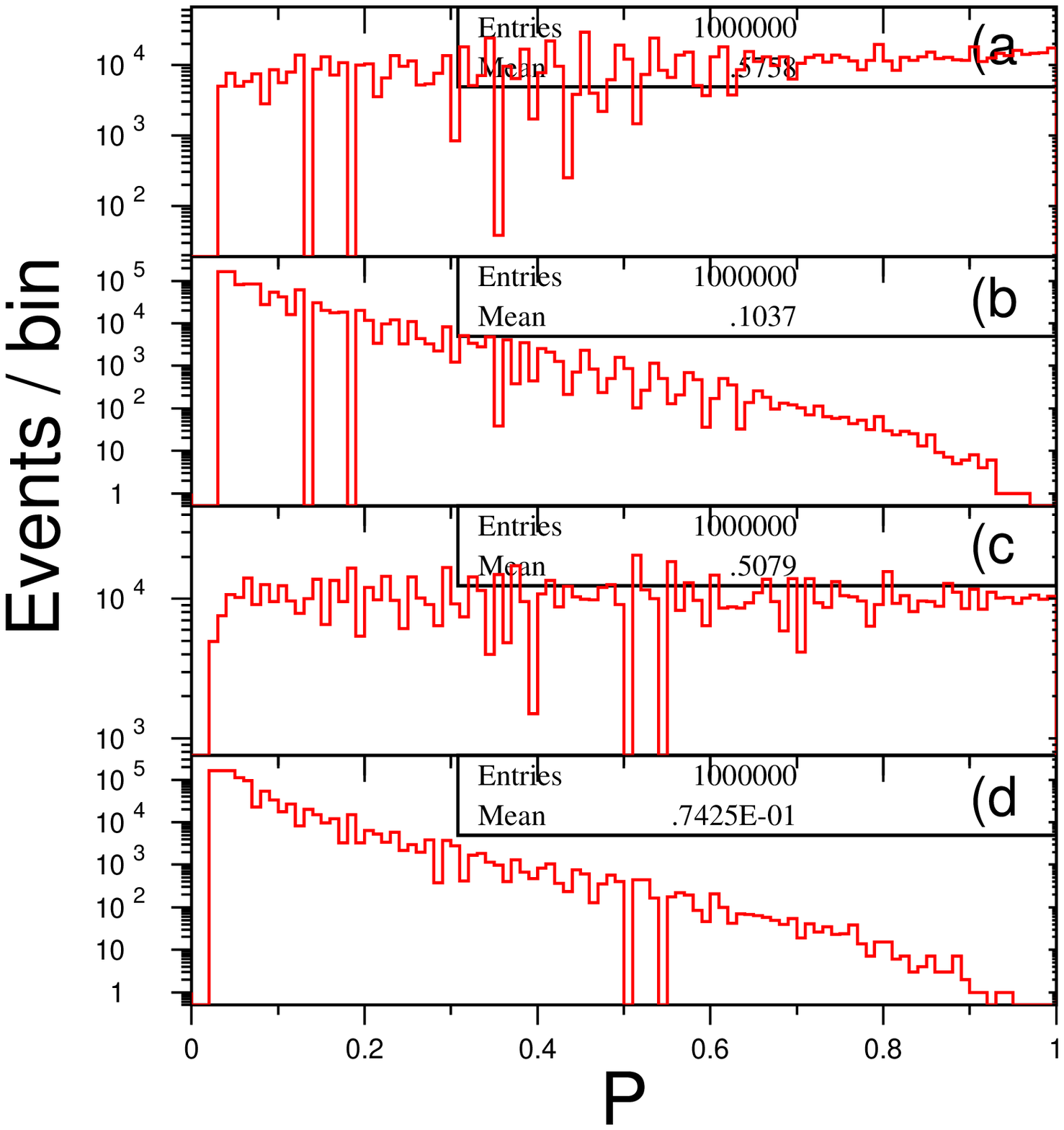}}
\end{tabular}
\caption{\label{met1} Distributions
for \PL\ in the signal + background (a) and background only (b) hypotheses 
and for \PT\ in the same hypotheses, (c) and (d), 
for a large number of trial experiments, starting from
analysis 1 alone and adding every time another analysis.
}
\end{center}
\end{figure}

\begin{figure}[hbtp]
\begin{center}
\begin{tabular}{cc}
\hspace{-1cm}
\mbox{\epsfysize=9.0cm \epsffile{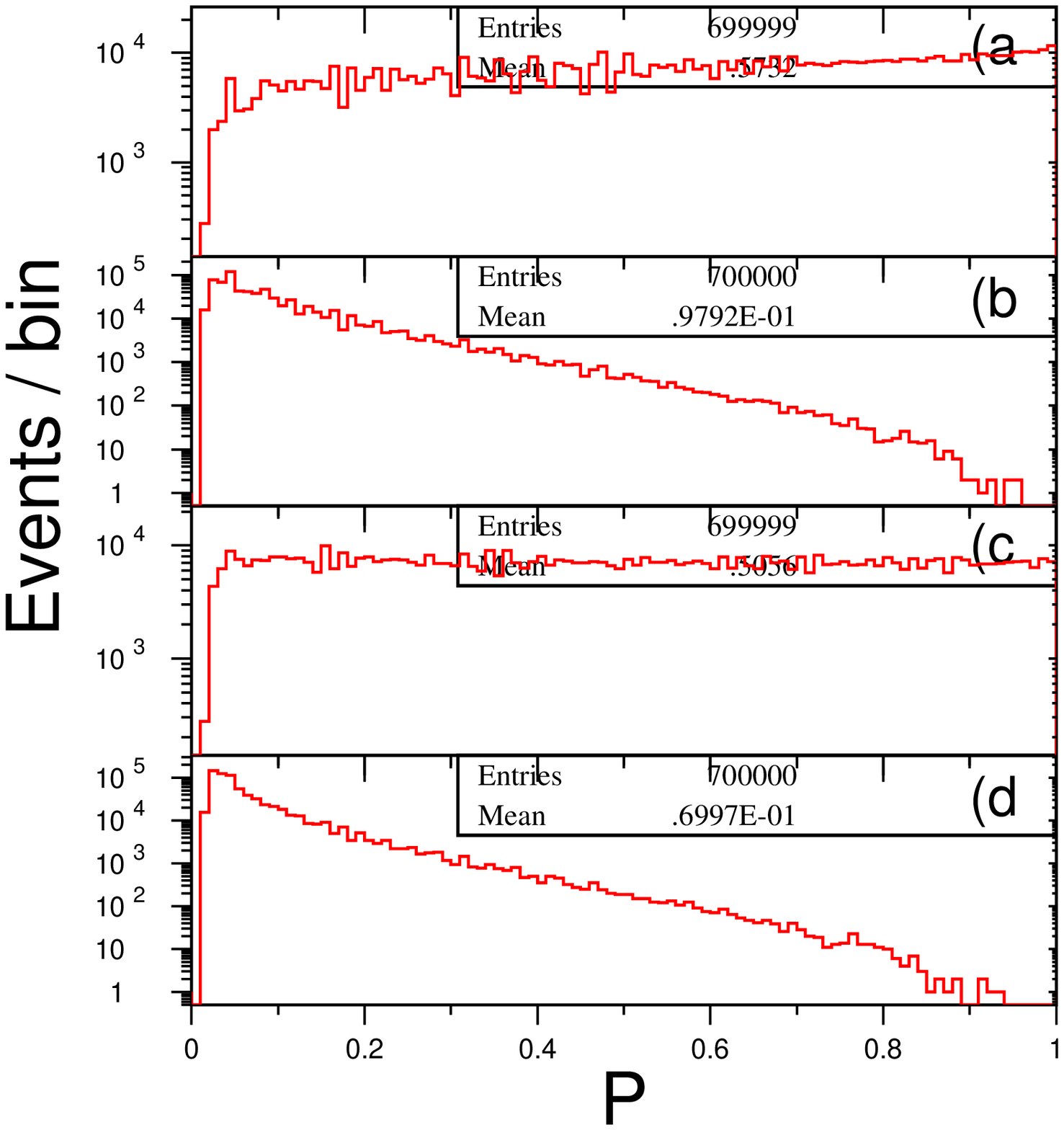}} &  
\mbox{\epsfysize=9.0cm \epsffile{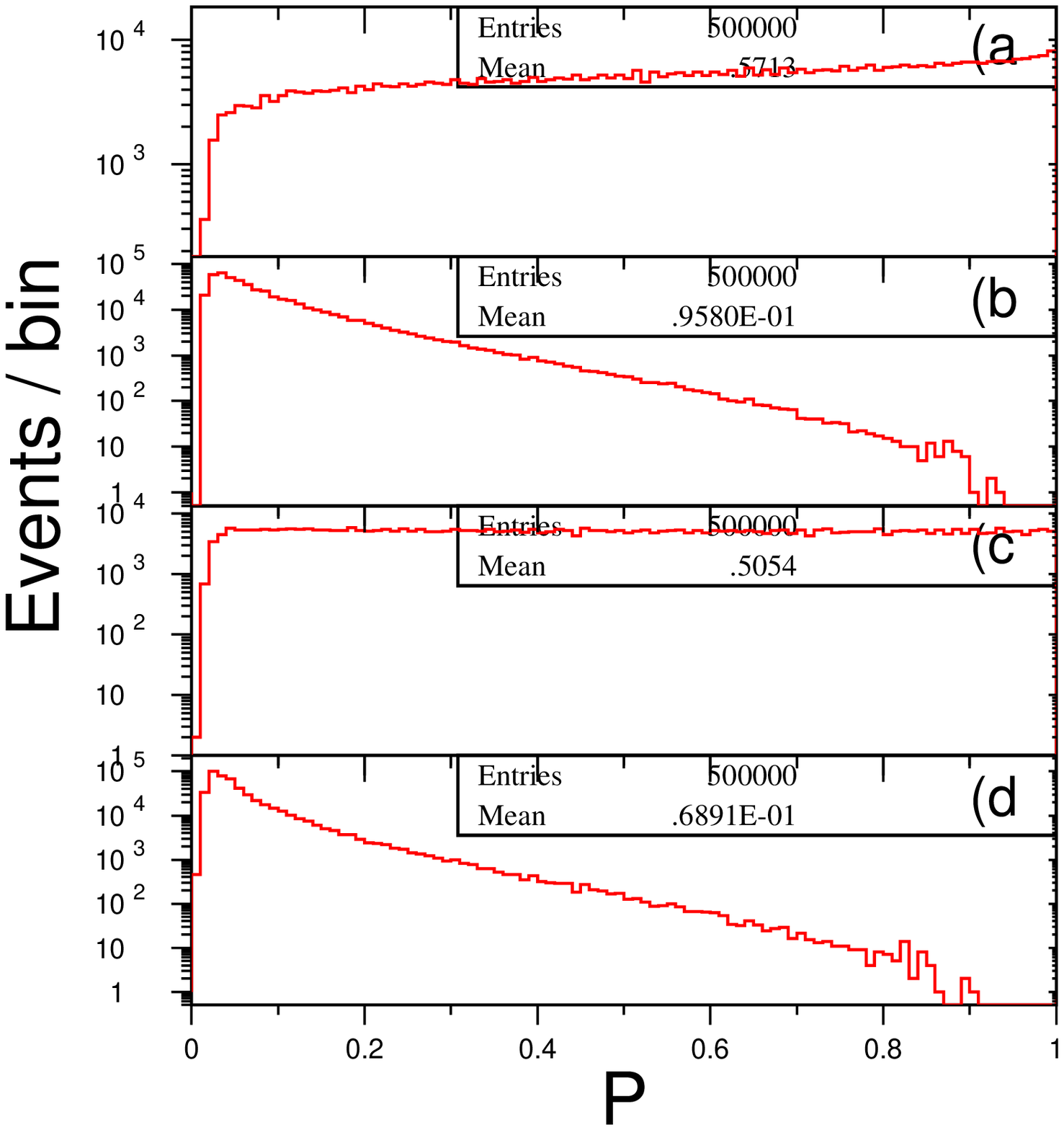}} \\
\hspace{-1cm}
\mbox{\epsfysize=9.0cm \epsffile{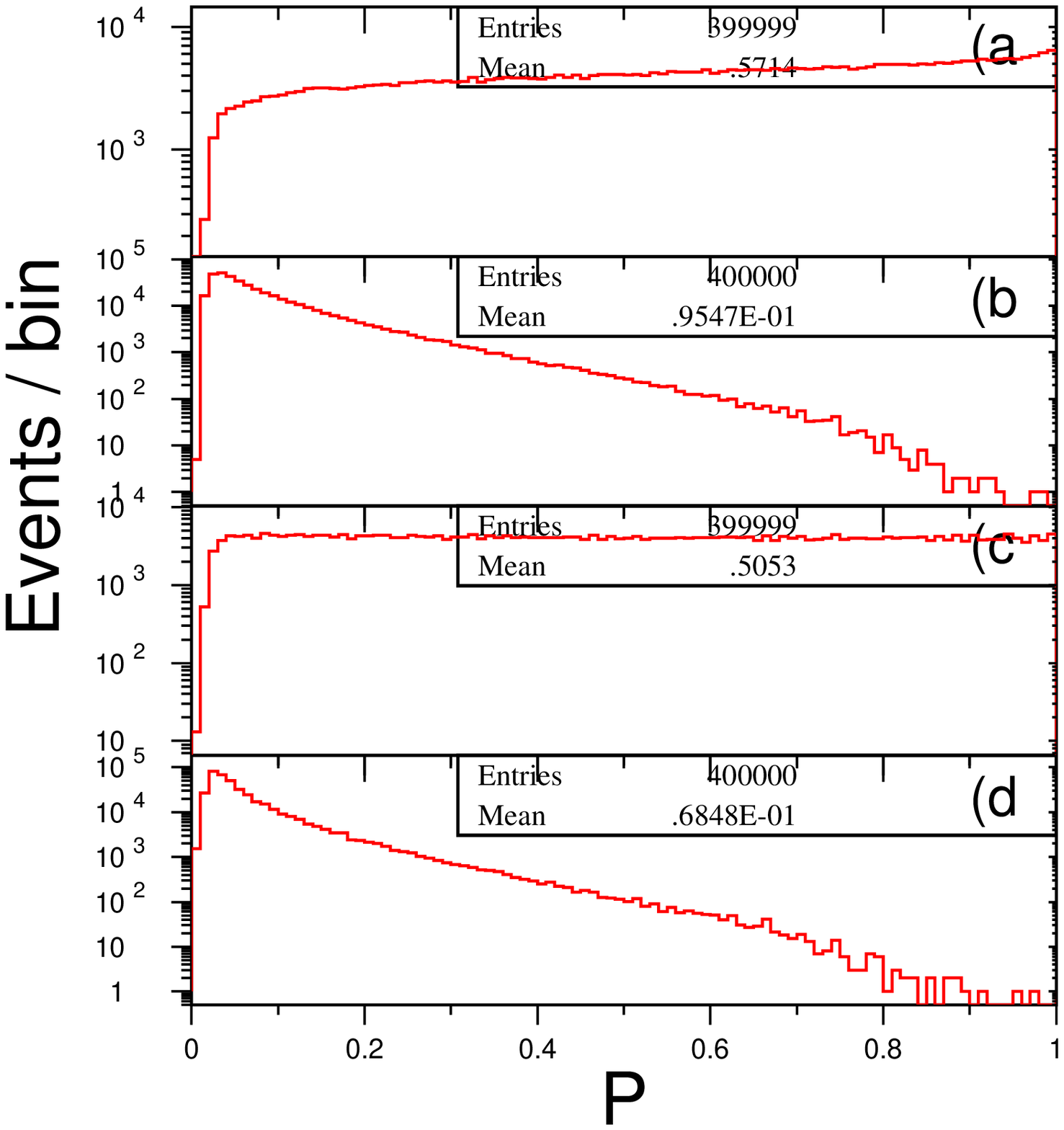}} &
\mbox{\epsfysize=9.0cm \epsffile{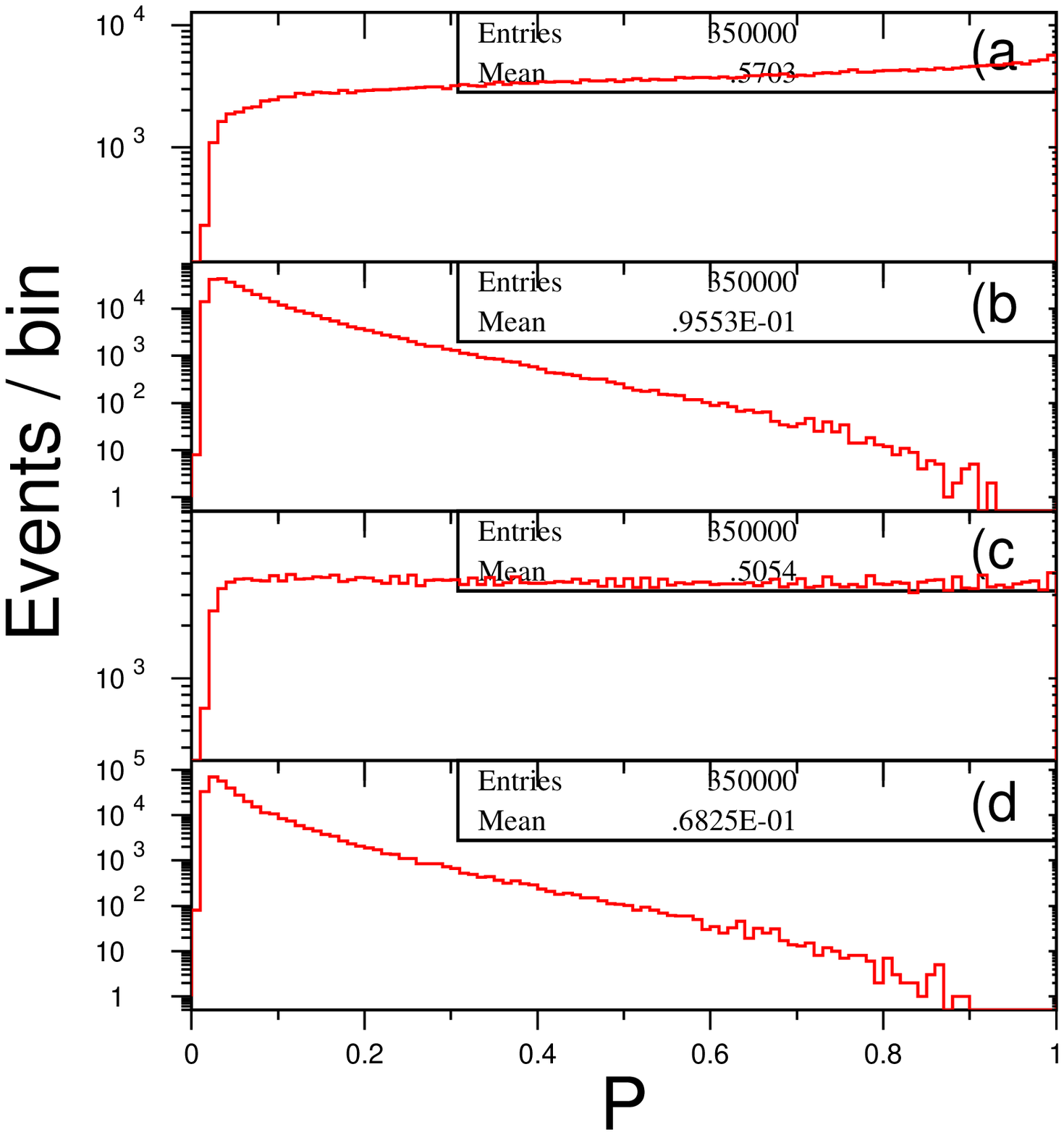}}
\end{tabular}
\caption{\label{met2} 
Distributions
for \PL\ in the signal + background (a) and background only (b) hypotheses 
and for \PT\ in the same hypotheses, (c) and (d), 
for a large number of trial experiments, starting from
the first five analyses and adding every time another analysis.
}
\end{center}
\end{figure}

We perform a set of MC experiments with the signal and background 
indicated for each bin, each time adding a new channel.
For each experiment we compute \PL\ as given by equation \ref{confi} for the 
two hypotheses:
background only and signal plus background. After correction to have
exact probability we get \PT, whose definition is described below, 
and we obtain the results shown in Figures \ref{met1} and \ref{met2}, 
where we show the distributions
for \PL\ and for \PT\ for a large number of trial experiments
(between 300 thousand and 1 million).

These Figures are ordered in
increasing number of included analyses.
The value of \PL\ is directly obtained from equation \ref{confi}
while the value \PT\ is obtained with Monte Carlo technique in
order to have the exact probability.
To obtain the exact probability, from MC experiments, for each 
value of our estimator \PL\ we should count the number of experiments 
 ($N_{exp}$) with 
a \PL\ for background+signal smaller than \PLV, where \PLV\ is the value
of the estimator obtained from the measurement.
Clearly, to properly take into account the information contained in the
experiment under study the background alone should have a \PL\ smaller
than \PLV. We should then divide by the fraction of experiments satisfying 
this last relation. Consequently we define \PT\ as:
$$\PT = \frac{\int_0^{\PLV} N_{exp}(\PL_{sig+back})
d\PL_{sig+back}}{\int_0^{\PLV} N_{exp}(\PL_{back}) d\PL_{back}}$$


\begin{table}
  \begin{center}
    \begin{tabular}{|c|c|c|}\hline
Last analysis included& Average \PL & Average \PT \\
\hline
  1 & $ 0.2535\pm .0001 $  & $ 0.2535\pm .0001$ \\    
  2 & $ 0.1299\pm .0001 $  & $ 0.1147\pm .0001$\\
  3 & $ 0.1119\pm .0001 $  & $ 0.0877\pm .0001$\\
  4 & $ 0.1028\pm .0001 $  & $ 0.0752\pm .0001$\\
  5 & $ 0.0978\pm .0001 $  & $ 0.0700\pm .0001$\\
  6 & $ 0.0956\pm .0001 $  & $ 0.0686\pm .0001$\\
  7 & $ 0.0956\pm .0001 $  & $ 0.0685\pm .0001$\\
  8 & $ 0.0954\pm .0001 $  & $ 0.0684\pm .0001$\\
  9 & $ 0.0955\pm .0002 $  & $ 0.0676\pm .0002$\\
\hline
    \end{tabular}
    \caption{Average \PL\ and  \PT\ for a large number of
 trial experiments, 
in the background only hypothesis. The first row includes only the
first analysis, while following rows include every time a new analysis
for a total of 9 analyses in the last row.
    \label{ratsb}}
  \end{center}
\end{table}

One indicator of the sensitivity of the experiment is, at this point, the
average \PT\ for experiments without signal, reported in table \ref{ratsb}
adding the bins of analysis one by one.

It is apparent that, when dealing only with statistical error
and performing background subtraction, the sensitivity of the
experiment is never degraded by the addition
of additional analyses regardless of the amount of background
that they contribute.
It can be seen from the average value of \PL\ and \PT\ that the gain is
remarkable if we include channels with signal/background ratio as low as 0.1, 
and is almost negligible for channels with signal/background ratio between
0.1 and 0.01.

In this study, even if not unambiguously demonstrated, the value
of \PL\ turned out to be always higher than the value of the true
probability; thus it seems impossible that, using \PL\ as the
true confidence level, optimistic limits were obtained.

\subsection{Statistical and systematic errors}

We repeat the same exercise this time taking into account
possible systematic errors on the signal efficiency and on the
background level and we want to check the stability of the method.

We perform a set of MC experiments with the signal and background 
indicated for each bin, each time adding a new channel.
As before for each experiment we compute \PL\ for the 
two hypotheses: background only, signal + background.

Now in addition we introduce a systematic uncertainty of
40\% both on signal efficiency and on the
background level applied when randomly generating the outcome
of the MC experiment. This systematic uncertainty is introduced
in each MC experiment according to a gaussian distribution
with mean equal to 40\%.

\begin{table}
  \begin{center}
    \begin{tabular}{|c|c|c|}\hline
Last analysis included& Average \PL & Average \PT \\
\hline
  1 & $ 0.2535\pm .0009$   &    $0.2880\pm .0009$\\
  2 & $ 0.1296\pm .0009$   &   $ 0.1693\pm .0009$\\
  3 & $ 0.1133\pm .0009$   &   $ 0.1428\pm .0009$\\
  4 & $ 0.1076\pm .0009$   &   $ 0.1307\pm .0009$\\
  5 & $ 0.1081\pm .0009$   &   $ 0.1437\pm .0009$\\
  6 & $ 0.1111\pm .0009$   &   $ 0.1451\pm .0009$\\
  7 & $ 0.1136\pm .0009$   &   $ 0.1614\pm .0009$\\
  8 & $ 0.1182\pm .0009$   &   $ 0.1530\pm .0009$\\
  9 & $ 0.1174\pm .0009$   &   $ 0.1642\pm .0009$\\
\hline
    \end{tabular}
    \caption{Average \PL\ and  \PT\ for 10000 trial experiments, 
in the background only hypothesis. The first row includes only the
first analysis, while following rows include every time a new analysis
for a total of 9 analyses in the last row. A systematic uncertainty of
40\% both on signal efficiency and on the
background level have been introduced.
    \label{ratsb1}}
  \end{center}
\end{table}

A systematic uncertainty of 40\% is really an extreme case but is
useful to check the results of the method.
From Table \ref{ratsb1} we realize that starting from the fifth analysis
(corresponding to a signal/background ratio of 0.25) 
the average sensitivity worsens. This is an important result since
it clarifies that according to the systematic uncertainty we
have to reject analyses with a signal/background ratio lower than
a certain value in order to reach the best sensitivity.

In Tables \ref{ratsb2} and \ref{ratsb3} we show the behaviour of the
Average \PL\ and  \PT\ when a systematic uncertainty of 20\% is
considered only for signal efficiency (Table \ref{ratsb2}) or
for background level (Table \ref{ratsb3}).

\begin{table}
  \begin{center}
    \begin{tabular}{|c|c|c|}\hline
Last analysis included& Average \PL & Average \PT \\
\hline
1&$0.2535\pm .0009$  &  $  0.2635\pm .0009$\\
2&$0.1291\pm .0009$  &  $  0.1261\pm .0009$\\
3&$0.1118\pm .0009$  &  $  0.1005\pm .0009$\\
4&$0.1039\pm .0009$  &  $  0.0825\pm .0009$\\
5&$0.0980\pm .0009$  &  $  0.0829\pm .0009$\\
6&$0.0965\pm .0009$  &  $  0.0838\pm .0009$\\
7&$0.0970\pm .0009$  &  $  0.0765\pm .0009$\\
8&$0.0961\pm .0009$  &  $  0.0809\pm .0009$\\
9&$0.0958\pm .0009$  &  $  0.0766\pm .0009$\\
\hline
    \end{tabular}
    \caption{Average \PL\ and  \PT\ for 10000 trial experiments, 
in the background only hypothesis. The first row includes only the
first analysis, while following rows include every time a new analysis
for a total of 9 analyses in the last row.
A systematic uncertainty of
20\% on signal efficiency have been introduced.
    \label{ratsb2}}
  \end{center}
%\end{table}

%\begin{table}
  \begin{center}
    \begin{tabular}{|c|c|c|}\hline
Last analysis included& Average \PL & Average \PT \\
\hline
1&$0.2539\pm .0009 $ &   $ 0.2597\pm .0009$\\
2&$0.1302\pm .0009 $ &   $ 0.1193\pm .0009$\\
3&$0.1130\pm .0009 $ &   $ 0.0913\pm .0009$\\
4&$0.1050\pm .0009 $ &   $ 0.0716\pm .0009$\\
5&$0.0997\pm .0009 $ &   $ 0.0745\pm .0009$\\
6&$0.1006\pm .0009 $ &   $ 0.0772\pm .0009$\\
7&$0.1007\pm .0009 $ &   $ 0.0773\pm .0009$\\
8&$0.1017\pm .0009 $ &   $ 0.0802\pm .0009$\\
9&$0.1017\pm .0009 $ &   $ 0.0773\pm .0009$\\
\hline
    \end{tabular}
    \caption{Average \PL\ and  \PT\ for 10000 trial experiments, 
in the background only hypothesis. The first row includes only the
first analysis, while following rows include every time a new analysis
for a total of 9 analyses in the last row. A systematic uncertainty of
20\% on the
background level have been introduced.
    \label{ratsb3}}
  \end{center}
\end{table}

\begin{figure}[hbtp]
\begin{center}
\mbox{\epsfysize=15.0cm \epsffile{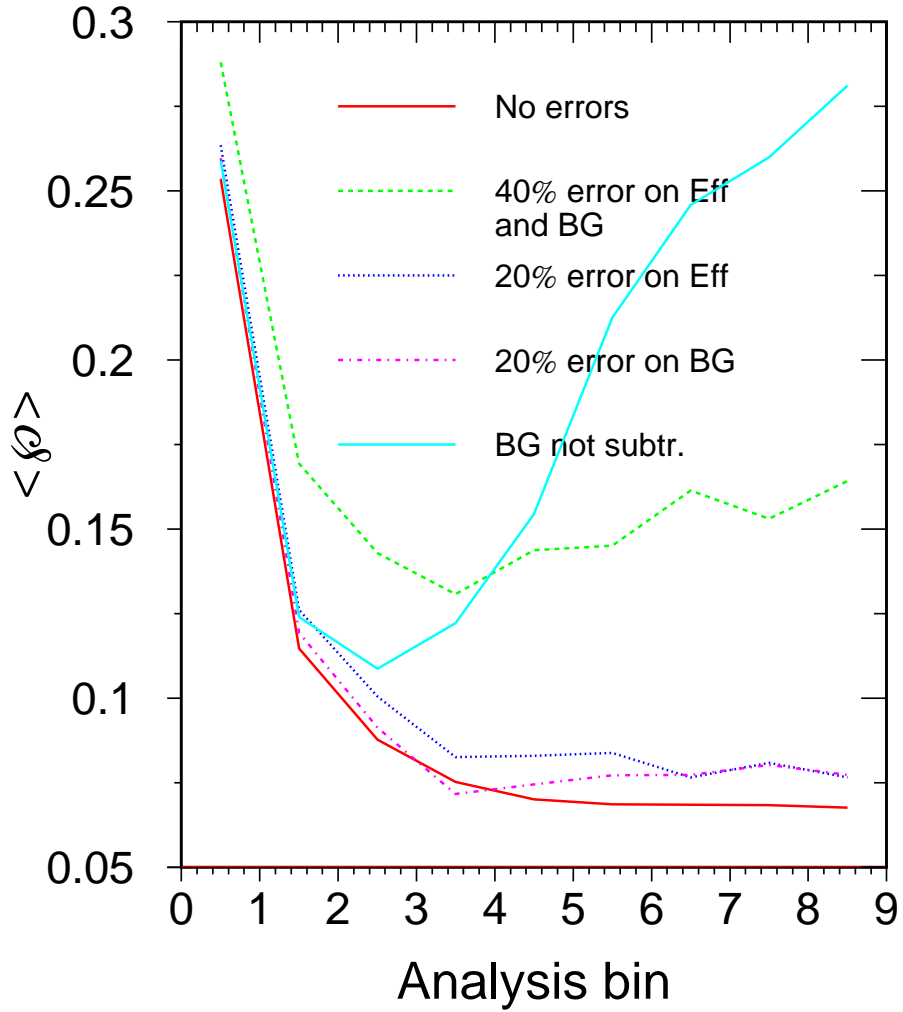}}
\caption{\label{ploterr} 
Average value of \PT\ as a function
of the number of analyses included. The five lines show the behaviour
of $<\PT >$ for different amplitudes of systematic errors
and in the case of no background subtraction.}
\end{center}
\end{figure}
\begin{table}
  \begin{center}
    \begin{tabular}{|c|c|c|}\hline
Last analysis included& Average \PL & Average ``\PT'' \\
\hline
1&$0.2532 \pm  .0009 $  &  $   .2588 \pm  .0009$\\
2&$0.1296 \pm  .0009 $  &  $   .1240 \pm  .0009$\\
3&$0.1111 \pm  .0009 $  &  $   .1087 \pm  .0009$\\
4&$0.1028 \pm  .0009 $  &  $   .1222 \pm  .0009$\\
5&$0.0982 \pm  .0009 $  &  $   .1544 \pm  .0009$\\
6&$0.0966 \pm  .0009 $  &  $   .2125 \pm  .0009$\\
7&$0.0971 \pm  .0009 $  &  $   .2458 \pm  .0009$\\
8&$0.0959 \pm  .0009 $  &  $   .2599 \pm  .0009$\\
9&$0.0951 \pm  .0009 $  &  $   .2811 \pm  .0009$\\
\hline
    \end{tabular}
    \caption{Average \PL\ and  ``\PT'' for 10000 trial experiments, 
in the background only hypothesis. The first row includes only the
first analysis, while following rows include every time a new analysis
for a total of 9 analyses in the last row.
No background subtraction has been applied to to compute ``\PT''.
    \label{ratsb5}}
  \end{center}
\end{table}

From Table \ref{ratsb2} we can not see any striking
worsening of the  sensitivity while in Table \ref{ratsb3}
it is rather significant the worsening starting from the sixth
analysis (signal/background ratio equal to 0.1). We can then conclude
that systematic errors on the background level are more important 
than those on the signal efficiency.

In Figure \ref{ploterr} we show the average value of \PT\ as a function
of the number of analyses included. In this Figure we compare
the behaviour of $<\PT >$ for different amplitudes of systematic errors.
In the same Figure we also show the behaviour of \PT\ in the case of 
no background subtraction.

This shows that even large systematic errors on the efficiency are
not significant while the same uncertainties on the background
play a role. In the case of huge systematic errors (40\%) the 
sensitivity degrades considerably including analyses with high level
of background. If one has to deal with such a situation it is
obviously better to reject analyses with too low signal to
background ratio (in our example between 0.1 and 0.25).

Finally in Table \ref{ratsb5} we show the values that
would be obtained if we neglected the existence of background
and we compared the outcome of the experiment
with signal only MC experiments.
If we did that the addition of high background analyses
would definitely deteriorate the sensitivity as can be clearly
seen from Figure \ref{ploterr}.

%%%%  no BACKGROUND subtraction

\section{Discoveries}

To estimate the analysis sensitivity for discoveries we proceed
exactly in the same way as previously shown when
optimizing the analysis sensitivity for exclusions.

The only difference is that we set the discovery \CL\ 
at the requested probability for the measurement to be consistent
with the background only hypothesis.

In Table \ref{discovery} we show the  probability to discover
our signal at different levels of significance (1\%, 0.1\% and 0.01\%)
as a function of the number of analyses included.

\begin{table}
  \begin{center}
    \begin{tabular}{|c|c|c|c|}\hline
Last analysis included&  Discovery at $\leq$ 0.01&  Discovery at $\leq$ 0.001
&  Discovery at $\leq$ 0.0001  \\
\hline
  1 &   0.4747 $\pm$ 0.2580 &
        0.2168 $\pm$ 0.1381 &
        0.0786 $\pm$ 0.0552 \\

  2 &   0.6088 $\pm$ 0.0779 &
        0.3492 $\pm$ 0.0414 &
        0.2183 $\pm$ 0.0499 \\

  3 &   0.6318 $\pm$ 0.0002 &
        0.3997 $\pm$ 0.0071 &
        0.2366 $\pm$ 0.0195 \\

  4 &   0.6292 $\pm$ 0.0021 &
        0.3988 $\pm$ 0.0027 &
        0.2312 $\pm$ 0.0089 \\

  5 &   0.6339 $\pm$ 0.0011 &
        0.3989 $\pm$ 0.0076 &
        0.2261 $\pm$ 0.0143 \\

  6 &   0.6326 $\pm$ 0.0030 &
        0.3991 $\pm$ 0.0092 &
        0.2063 $\pm$ 0.0245 \\

  7 &   0.6338 $\pm$ 0.0031 &
        0.3973 $\pm$ 0.0090 &
        0.2453 $\pm$ 0.0246 \\

  8 &   0.6296 $\pm$ 0.0030 &
        0.3877 $\pm$ 0.0144 &
        0.2237 $\pm$ 0.0120 \\

  9 &   0.6345 $\pm$ 0.0030 &
        0.4006 $\pm$ 0.0101 &
        0.2254 $\pm$ 0.0209 \\

% 10 &   0.6352 $\pm$ 0.0051 &
%        0.4017 $\pm$ 0.0102 &
%        0.2005 $\pm$ 0.0216 \\

\hline
    \end{tabular}
    \caption{Discovery probability at different \CL.
The first row includes only the
first analysis, while following rows include every time a new analysis
for a total of 9 analyses in the last row.
    \label{discovery}}
  \end{center}
\end{table}

From this Table we can see that the gain due to the inclusion
of low purity analyses is much less than for the case of exclusion
as shown in Table \ref{ratsb}. There we had a remarkable gain
including analyses with signal/background ratio as low as 0.1,
while in Table \ref{discovery} we have a remarkable gain only
including analyses with signal/background ratio not lower than 1.

\section{Measurements and other issues}

%This section is more preliminary than the others
%because we did not have time yet.\\
Using the same estimator described before is possible to
directly measure the rate of a given process. It can easily be seen
that the optimization method here proposed also improves
the errors on the measurement.

Another issue which follows naturally from this optimization
method is the combination of results from different
experiments. From what we have described it should be clear
that different experiments can be treated on the same footing just
as different analyses with their own efficiencies, backgrounds
and systematic errors.

\section{Conclusions}

We propose a new method, 
which can be applied to searches and
to any cross section measurement.

Using a large variety  of estimators we 
are able to devise an intrinsically optimized analysis
which is the same for discovery (and relative cross section
measurement) and for exclusion.

Using this method the sensitivity of the analysis
is improved compared to most of previously used methods.

%\section*{Acknowledgements}

%We wish to express our gratitude to J. Branson and R. Faccini
%for kindly reading this note.
%

%%%%%%%%%%%%%%%%%%%%%%%%%%%%%%%%%%%%%%%%%%%%%%%%%%%%%%%%%%%%%%%%%%%%%%%%%%%%%%%
% Bibliography
%%%%%%%%%%%%%%%%%%%%%%%%%%%%%%%%%%%%%%%%%%%%%%%%%%%%%%%%%%%%%%%%%%%%%%%%%%%%%%
%
% Style file to use with mcite.
% Use l3style with just cite.
\bibliographystyle{l3stylem}
\bibliography{%
l3pubs,%
mybibnew,%
tesi}

\end{document}